\documentclass[10pt, conference]{IEEEtran}

\usepackage[hidelinks]{hyperref}

\IEEEoverridecommandlockouts

\newcommand{\remove}[1]{}

\usepackage{ifpdf}
\usepackage{graphicx}
\ifpdf
 \PassOptionsToPackage{pdftex}{graphicx}
 \usepackage{epstopdf}
\else
 \PassOptionsToPackage{dvips}{graphicx}
\fi

\usepackage{amsmath}
\usepackage{algorithm}
\usepackage{flushend}

\begin{document}
%
\title{On the Relevance of Wait-free Coordination Algorithms in Shared-Memory HPC:\\The Global Virtual Time Case}

\author{

\IEEEauthorblockN{Alessandro Pellegrini, Francesco Quaglia}
\IEEEauthorblockA{DIAG, Sapienza Universit\`a di Roma
}
}

\maketitle

\begin{abstract}
\boldmath
High-performance computing on shared-memory/multi-core architectures could suffer from non-negligible
performance bottlenecks due to coordination algorithms, which are nevertheless necessary to
ensure the overall correctness and/or to support the execution 
of housekeeping operations, e.g. to recover computing resources
(e.g., memory). Although more complex in design/development,
a paradigm switch from classical coordination algorithms to 
wait-free ones could significantly boost the performance of HPC applications. 

In this paper we explore the relevance of this paradigm shift in shared-memory architectures, by focusing
on the context of Parallel Discrete Event Simulation, where the
Global Virtual Time (GVT) represents a fundamental coordination algorithm.
It allows to compute the lower bound on the value 
of the logical time passed through by all the entities
participating in a parallel/distributed computation. 
Hence it can be used to discriminate what events belong
to the past history of the computation---thus
being considered as committed---and
allowing for memory recovery (e.g. of obsolete 
logs that were taken in order to support state recoverability) and non-revokable operations (e.g. I/O). 

We compare the reference (blocking) algorithm for shared memory,
the one proposed by 
by Fujimoto and Hybinette \cite{Fuj97}, with an innovative wait-free implementation,
emphasizing on what design choices must be made to enforce this paradigm shift,
and what are the performance implications of removing critical sections
in coordination algorithms.

 \end{abstract}

\IEEEpeerreviewmaketitle

\section{Introduction}
\label{intoduction}

Developing high performance applications, e.g., systems
able to provide faster than real-time results, is a core objective for
differentiated contexts, such as when
virtual and real worlds interact for either training purposes (see, e.g., \cite{training}), or for system prediction/audit in scenarios where components are still under design/development
(see, e.g., \cite{cptvt}).

Independently of the application domain, most parallel/distributed applications
rely (at some level) on coordination algorithms, which are fundamental to, e.g., keep track of the evolution
of computation for termination detection, for load balancing, or for fine-tuning of
runtime parameters. The advent of shared-memory/multi-core architectures has made
parallel applications proliferate in a wide number of contexts, proving the importance
of the field, and requiring always more accurate solutions for performance enhancement.
Nevertheless, the traditional approach to realize coordination algorithms is to
rely on locking primitives. It has been shown in literature that lock-based programming
can suffer from several problems, like deadlocks, livelocks, convoying, priority inversions \cite{Cha07},
and in case of very high contention it can provide a significant slowdown of the whole parallel
application.

A more recent trend is to rely on non-blocking algorithms \cite{Her91}, which exploit very fine-grained
synchronization (i.e., at the level of single hardware instructions) to avoid the execution of traditional ciritical
sections while ensuring the correctness of the outcome. 
Avoiding mutual exclusion has been considered a benefit since the early 1970's \cite{Eas72}.
Lamport \cite{Lam77} gave the first non-blocking algorithm for the problem of a single-writer/multiple-reader
shared variable.
Herlihy \cite{Her91} proved that for non-blocking implementations
of most interesting data types, a synchronization
primitive that is universal, in conjunction with reads
and writes, is both necessary and sufficient. A universal
primitive is one that can solve the consensus problem \cite{Fis85} for
any number of processes. Universal primitives commonly supported by out-of-the-shelf
hardware architectures are, e.g., Compare\&Swap ({\tt CAS}) and Load-Link/Store-Conditional ({\tt LL/SC}).

Although several fundamental data structures have already been made accessible by non-blocking
algorithms when running on multi-core systems (like, e.g., stacks \cite{Tre86}, queues \cite{Kog11}, linked lists \cite{Har01}, or doubly linked lists \cite{Mar09}),
the complexity related to the design/development of non-blocking algorithms is slowing the pace
at which they are being adopted in high performance computing, although some fields (like, e.g.,
Software Transactional Memories \cite{Sha95}) are making a larger use of them. The goal of this
paper is to attract the attention of the research community towards this emerging field.

As a case study, we have taken Discrete Event Simulation (DES), for which
a classical technique to achieve high performance is Parallel-DES (PDES).
It is based on partitioning the simulation
model into several distinct objects, also known as Logical
Processes (LPs) \cite{Fuj90},
which concurrently execute simulation events, thus allowing for exploitation of parallelism in the underlying
hardware architecture.
The main problem in designing/developing this type of
simulation platforms is synchronization, the goal of which
is to ensure causally-consistent (e.g. timestamp-ordered) execution of simulation
events at each concurrent LP.
In literature,
several synchronization protocols
have been proposed, among which the optimism-oriented ones, such as the Time Warp protocol
\cite{Jef85}, are recognized to be highly promising.

In Time Warp, block-until-safe policies for event processing at the LPs are avoided,
thus allowing speculative computation, which is reflected into great exploitation of parallelism.
 At the same time, causal consistency is guaranteed
through rollback/recovery techniques,
which restore the system to a correct state upon the a-posteriori
detection of consistency violations. These are originated when $LP_a$ schedules a new event destined
to $LP_b$ having a timestamp lower than the one of some event already speculatively processed by $LP_b$.
In case this occurs, the rollback of $LP_b$ might also require undoing the send operation of events that were produced by $LP_b$ during the rolled back portion of the computation. This is usually achieved via so called anti-messages (carrying anti-events), which are aimed at annihilating the originally-sent events, thus possibly causing a cascading rollback across chains of LPs.

A core abstraction underlying Time Warp-based platforms is Global Virtual Time (GVT),
which is defined as the smallest timestamp among
those of events (or anti-events) that are still unprocessed, or that are currently being processed.
Since no LP can ever rollback to simulation time preceding the
GVT \cite{Jef85}, its value indicates the commitment horizon of
the speculative simulation run. It is used both to execute actions that cannot be
subject to rollback, such as displaying/inspecting intermediate simulation
results (see, e.g., \cite{Ant13,Cuc07}), and for recovering memory (see, e.g., \cite{Das97b}).  Specifically, events with
timestamp lower than GVT will never need to be re-executed after a
rollback, therefore they can be discarded. The same happens to obsolete state information, if any,
maintained to support recoverability of the LP state.  The action of
recovering memory after GVT calculation is typically referred to as
{\em fossil collection}.

Computing the GVT value, e.g., on a periodic basis, requires the Time Warp platform to implement some GVT coordination algorithm, and different algorithms have been devised (see, e.g., \cite{Mattern93,Bau05}) depending on the specific features of the underlying platform (e.g. shared vs distributed memory) hosting the Time Warp system. For shared memory platforms, the reference GVT algorithm is the one provided by Fujimoto and Hybinette in \cite{Fuj97}. This proposal exploits the so called {\em observability} property, commonly matched by shared memory implementations of Time Warp, in order to provide a GVT protocol that does not require any message acknowledgment, which is instead employed by GVT protocols suited for distributed memory versions of Time Warp (see, e.g., \cite{tel,Lin89,LaiY87}).

Essentially, in an observable Time Warp system, the send operation of any message/anti-message leads to the direct incorporation of the sent information into the message-queuing data structure of the destination, which removes the notion of in-transit message.
This feature is exploited in \cite{Fuj97} to devise a GVT protocol relying on a single phase in which, once aware of a new GVT computation request, each process simply (A) keeps track of the minimum timestamp of any message/anti-message it sent out (towards any other process) and (B) after incorporating into its event-queue any message/anti-message detected as incoming in its message-queueing data structure,
it contributes to the GVT computation by writing its so called {\em local minimum}, namely the minimum across the timestamps of its sent out and incoming messages/anti-messages, into a proper memory location. The last process that ends the above tasks is also in charge of computing the {\em global minimum} (namely the GVT value to be adopted) across all the local minima.

In this approach, the start of the GVT computation phase takes place by atomically setting a GVT-flag to the value $N$, which corresponds to the the number of participating processes. The GVT-flag is then decreased by one each time a process ends the activities in the above task B, and is atomically checked to have reached the value zero in order to trigger the computation of the global minimum value. Overall, task B is executed within a critical section (which possibly includes the actual computation of the global minimum), namely in a sequentialized fashion, in order to guarantee correctness and progress of the GVT protocol. However, such a critical section  may represent an impairment to scalability, thus potentially hampering performance in contexts where the Time Warp system is deployed on top of machines with non-minimum CPU-core counts.

To assess the effects on performance of wait-free coordination, in this article we compare the execution
time of the traditional GVT coordination algorithm by Fujimoto and Hybinette \cite{Fuj97} with a new implementation
which does not require any critical section, hence standing as a {\em wait-free algorithm} \cite{Her91}.
Our proposal is based on memory atomic operations, namely {\tt CAS}, which are used to
keep track of the advancement of any process within different phases of the GVT protocol. As opposed to \cite{Fuj97}, our GVT protocol requires multiple phases (rather than one), but given the wait-free nature of any task carried out in any of its phases, it can reduce the actual cost (and also the latency) for computing the new GVT value especially for deploys on non-minimal scale multi-core systems.
Overall, the wait-free GVT protocol is aimed at better coping with scalability aspects of Time Warp platforms to be run on top of multi/many-core shared-memory platforms. Also, its wait-free nature allows to better cope with contexts where the computing platform can be shared across Time Warp processes and other kinds of workload that may interfere on CPU usage, thus possibly stretching the time-span of the critical section required in the protocol in \cite{Fuj97} (e.g. in case a Time Warp thread currently running the critical section is context-switched off the CPU). With our proposal, no wait-phase is induced across Time Warp threads in case some or more of them are context-switched off the CPU while GVT computation is in progress, in fact all the other threads can continue processing simulation events.

We have implemented our GVT protocol into the open source {\sf ROOT-Sim} speculative simulation platform \cite{rootsim}, exactly based on the Time Warp paradigm, and we have also performed tests assessing the effectiveness of the algorithm when running this platform on top of a 64-bit NUMA machine equipped with 32 CPU-cores using a version of the well known PHOLD benchmark \cite{Fuj90b} as the application test-bed.

The remainder of this paper is organized as follows. In Section \ref{related} we discuss related work. The new GVT protocol is presented in Section \ref{protocol}. Experimental results are provided in Section \ref{data}.

\section{Related Work}
\label{related}

In literature, several wait-free implementations of data structures have been proposed.
The seminal work in \cite{Tre86}---which was conceived as a motivation for the introduction
of the {\tt CAS} primitive in the IBM 370 in 1970 to reduce the use of spinlocks---presents
a practical implementation of a stack (realized as a linked list) where the {\tt CAS}
operation is used to update the top of the stack's pointer, to support atomic insertion
and deletion of nodes at the head of the list.

In \cite{Har01}, 
a non-blocking linked list is realized by differentiating the number of {\tt CAS} operations which are
required for implementing the \textit{insert} and the \textit{delete} operations. Specifically,
for a thread/process to insert a new node into the list, it can scan the list to find the proper place
where the new node should be inserted (according to some ordering, e.g., increasing key value)
and using the {\tt CAS} to alter the pointer of the previous node. Correctness in ensured by
the fact that a modification to the previous node will make the {\tt CAS} fail, forcing the
thread/process to retry the operation (thus taking into account the updated view of the list).
A delete operation is split over two different {\tt CAS} instructions, the first being used
to mark a pointer as ``logically deleted'', and the second one being used to ``physically
delete the node'' by making the previous node point to the next one.
A conceptually similar proposal in \cite{Mar09} allows to realize wait-free doubly-linked lists.

In the context of PDES, the work in \cite{Pell12} allows a LP to overcome the limitation
of having a completely disjoint state by supporting concurrent accesses on global variables.
By relying on static software instrumentation \cite{Pell13b}, each global variable is transformed
to a multi-version list, which is then accessed concurrently using a modified version of the
algorithm in \cite{Har01}. 

On the side of coordination algorithms, GVT protocols can be barely divided in two categories,
depending in whether they can cope with distributed memory systems, or require tightly coupling of nodes in a shared memory platform.

As for the first category, several algorithms have been proposed, which have been based on explicit message acknowledgment schemes \cite{tel,Lin89,LaiY87} in order to determine which messages (or anti-message) are still in transit and which processes are responsible for keeping into account the timestamps of in-transit messages in the computation of the new GVT value. Some of these algorithms (see, e.g., \cite{tel,LaiY87}) opt for acknowledging each individual message, which reduces the time interval along which a message can result as still in-transit. On the other hand, other approaches (see, e.g., \cite{Lin89}) opt for acknowledging batches of messages (rather than individual ones) which allows for reducing the message-overhead by the protocol, but stretches the interval of time along which a message still results in-transit (although being potentially already processed at the destination). This, in its turn, leads to  worsening the approximation provided by the algorithms on the actual GVT, given that ``obsolete" timestamps might be still considered in the global reduction  computing the new GVT
value to be adopted.

An approach where explicit message acknowledgments are not required has been provided in \cite{Mattern93}. In this solution, messages are associated with kind of ``phases" (represented by different messages coloring schemes) so that it is possible for the processes in the system to determine whether the timestamp of any message (or anti-message) needs to be accounted for in the current GVT computation. However, the protocol requires control messages to set-up the start of new protocol phases.

The need for both control messages and acknowledgments is removed by the proposal in \cite{Bau05}, which has been tailored to distributed memory clusters where specific bounds can be assumed on the message delivery transfer across the nodes in the system, and the clocks of the different machines can be assumed to be (perfectly) synchronized. In this proposal, new execution phases of the GVT protocol are triggered by specific timeouts, which occur in synchronized way across all the nodes in the system, thus giving  rise to the scenario where all the nodes observe the start of the GVT protocol at the same identical time instant, and are able to determine which messages (or anti-messages) can be still in transit since the start of the current GVT computation, given the knowledge on the upper bound delivery delay. These messages are accounted for by the sender in the global reduction associated with the newly computed GVT value.

In this paper, we explicitly target shared memory Time Warp systems. Hence our analysis stands on an orthogonal
setting for the execution of the coordination algorithm.

As pointed out before, for the case of tightly coupled shared memory systems, the reference GVT protocol is the one by Fujimoto and Hybinette \cite{Fuj97}. This protocol requires the Time Warp system to be observable, a property which we have already pointed out, and which expresses that no message (or anti-message) can ever be in-transit, given that the corresponding send operation leads to directly incorporating the message into the recipient message-queue. In this protocol, the start of the GVT computation phase is instantaneously visible to all the processes, given that it simply requires setting a proper flag into shared memory to the number of participating processes. However,
as we already pointed out, the computation of the local minimum at each process and the decrease of the counter in order to indicate that the contribution by the process has been made available, are executed within a critical section, which may represent a major impairment to scalability and resilience to interfering external workload (which may impact the duration of the critical section). 
We consider this work as a baseline for the comparison with out wait-free implementation, which avoids any critical section by trading-off in a completely different way synchronization costs and the number of phases required to compute the updated GVT value.

Similar considerations apply to the work in \cite{XiaoGUC95}, which presents a GVT protocol for observable shared memory Time Warp systems (although observability was formally defined later in literature by \cite{Fuj97}). In particular, this work is based on a critical section that is used to atomically update the entries of an array of elements, with size equal to the number of participating processes/threads. This makes the protocol non wait-free.

  \remove{

 the different processes are not meant to check whether GVT computation is ongoing simultaneously. As a consequence, it is possible that some process, which is already aware of the new GVT computation phase, computes its local minimum  (namely the mimimum timestamp of all the messages incorporated into the message-queues of the LPs it is managing) while some other process is not. The latter might send some message/anti-message to the former, whose timestamp is not accounted by the destination in the GVT computation. To overcome this problem, as soon as a process becomes aware of a new GVT computation phase, it needs to store the minimum value of the timestamp of any message it sent out towards other processes during the current GVT phase. The local miminum supplied by any process, which is ultimately used for computing  the global minimum representing the GVT, is evaluated by the process as the minimum of any message it sent out and any message still to be processes, which is already incorporated into its message-queue. In this proposal, any acknowledgement message is avoided, thanks to the atomic incorporation of any sent message (or anti-message) into the message-queue of the destination process.

Compared to the protocol in \cite{}, our solution does not require that

}



\section{The Wait-Free GVT Protocol}
\label{protocol}


\begin{figure}
\centering
\includegraphics[height=1.0in]{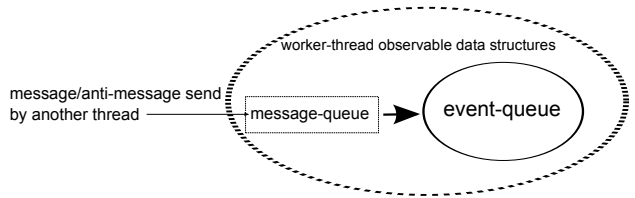}
\caption{Data structures and send operations in observable Time Warp systems (the unique event-queue is the logical collection of the corresponding event-queues of the LPs managed by the worker-thread).}
\label{observability-scheme}
\end{figure}

\subsection{Protocol Overview}

In classical implementations of Time Warp, each worker-thread running within the simulation platform is in charge of managing a set of LPs. Particularly, it is in charge of handling the event-queues of these LPs, which are used to keep all the events that have been scheduled for them.
In shared memory Time Warp versions (see, e.g., \cite{DasFPAH94,Vit12b}) the worker-thread is also associated with an input messaging data structure (the message-queue), where messages/anti-messages incoming from other worker-threads are directly buffered right upon the corresponding send operation execution (see Figure \ref{observability-scheme}). The message-queue and the event-queue are directly accessible by the corresponding worker-thread so that it can ``observe" at any time the value of the timestamps of any message/anti-message existing in the system, which is destined to the LPs it is managing. Hence no message/anti-message is ever in-flight across worker-threads, hence being not accessible (in terms of ability to read its timestamp) by the destination worker-thread.

Given that, in this kind of organization, GVT represents the global minimum value (across all the worker-threads) of the timestamps of messages/anti-messages that are either into the message-queue or that have already been incorporated into the event-queue (in fact no in-transit message exists), building a GVT protocol actually means determining the right moment for the worker-thread to look at its data structures and to compute its local minimum, which will be then used for the calculation of the global minimum.
With no loss of generality, in our approach the local minimum will be computed after having incorporated the already present incoming messages/anti-messages into the event-queue, meaning that if an anti-message cancels a specific event, then the timestamps of both the canceled event and the anti-message will no longer have to be accounted for. This complies with observability operations as described in
\cite{Fuj97}, while simplifying  the computation procedure, given that the local minimum will correspond to the minimum value of the timestamps of events kept by the event-queue. Also, the incorporation of any message/anti-message is meant to leave the event-queue in a causally consistent state, with the meaning that if a message/anti-message incorporation leads some LP to be flagged for rollback,
then
the event-queue is refilled with the already processed events that need to be reprocessed after the LP rollback is finalized.
This again complies with the specification of observability operations provided in
\cite{Fuj97}.

\begin{figure}
\centering
\includegraphics[height=1.2in]{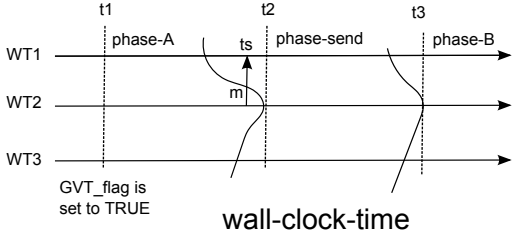}
\caption{GVT protocol-phases example.}
\label{phases-example}
\end{figure}

In our approach, we determine the right moment(s) for a worker-thread to look at its data structures and compute the local minimum by having the all the worker-threads participating in the GVT computation to pass through a set of different phases. No successive phase can be entered by any worker-thread unless all the other worker-threads have already executed the previous protocol phase. An example picture for this type of behavior is shown in Figure \ref{phases-example}, where we represent the start of the GVT protocol as the atomic set of a special GVT-flag to the value $TRUE$, operation that occurs (at time $t_1$ in the example) as instantaneously visible to all the worker-threads, given the shared memory (and cache coherent) nature of our target platform. On the other hand, the conclusion of the different phases on different threads can occur at different time instants of wall-clock-time, as explicitly shown in the picture (although the actual system-wide end of the phase, and the begin of the successive phase, corresponds to the latest wall-clock-time instant where the end occurs on some process).

Our approach is based on partitioning the GVT protocol in a sequence of phases according to which each worker thread $WT_i$ computes its local minimum two times, hence determining two values $min_i^A$ and $min_i^B$. The actual local minimum provided by $WT_i$ for the computation of the global minimum value  will then result as $min(min_i^A, min_i^B)$. Between the two computations of $min_i^*$, whose phases are referred to as {\sf phase-A} and {\sf phase-B}, we interpose an additional phase, marked as {\sf phase-send}
(see Figure \ref{phases-example}). The {\sf phase-send} is such that each worker-thread $WT_i$ is requested to process at least one pending event (destined to some LP it is managing), if any, and to send newly scheduled events produced during such processing phase towards the destination worker-threads. The {\sf phase-send} starts in our approach right after all the worker-threads ended their tasks related to {\sf phase-A}. For the example in Figure \ref{phases-example} this occurs at wall-clock-time $t_2$. We note that when the last one of the send operations by $WT_i$ is performed while being in the {\sf phase-send}, any other message/anti-message previously produced by $WT_i$ (destined to whichever worker-thread) is guaranteed to be already incorporated into the destination data structures (namely the message-queue of the destination worker-thread), given the intrinsic sequential nature of the activities at worker-thread $WT_i$ and system observability.

At this point, indicating with $MINTS_i$ the minimum timestamp of any message/anti-message
 sent by $WT_i$ up to the end of the {\sf phase-send}, we have the following two possibilities:

\begin{itemize}
\item[(A)] $min_i^A \leq MINTS_i$, in this case $min_i^A$ incorporates the lower bound on the logical time value that can be affected by any activity possibly occurring (or already occurred) at $WT_i$ up to the end of the {\sf phase-send}.
\item[(B)] $min_i^A > MINTS_i$, in this case $min_i^A$ does not represent the lower bound on the the logical time value that can be affected by some activity occurred at $WT_i$.
\end{itemize}

However, given that any worker-thread $WT_j$ recomputes $min_j^B$ after  {\sf phase-send} is already over, all the messages/anti-messages sent by any worker-thread to $WT_j$ up to the end of  {\sf phase-send}, and hence up to the end of {\sf phase-A}, have been already incorporated into the data structures handled by $WT_j$ (some of them might have been already processed, thus already belonging to the past of the computation). Hence, $min_j^B$ represents the lower bound on the logical time value that can be affected by $WT_j$ when also considering incoming information after the {\sf phase-send} is over. Therefore, $min(min_i^A, min_i^B)$ is the absolute lower bound on the logical time value that can be affected by the generic worker-thread $WT_i$  after the {\sf send-phase} is over.
By having each worker-thread $WT_i$ writing $min(min_i^A, min_i^B)$ into a proper memory location, and then computing the absolute minimum across all the values kept by these locations we determine the value for the GVT. Such a computation is realized in our scheme via an additional phase, occurring after all the worker-threads have posted their local minimum into their associated memory locations (e.g. the entries of a shared array, each one associated with a specific worker-thread).

We note that, if the worker-threads where perfectly synchronized, thus computing their $min_*^A$ values at the same identical time instant, then GVT  would simply correspond to the global minimum across these values, given system observability. However, to avoid thread synchronization, $min_*^A$ values are computed at different time instants, hence some message/anti-message might have not yet been incorporated into the message-queue of the recipient before the sender computes it $min_*^A$ value. An example is shown in Figure \ref{phases-example}, where a message $m$ is sent at wall-clock-time $t_s$ from $WT_2$ to $WT_1$ after $WT_1$ already computed $min_i^A$, but before $WT_2$ computes $min_2^A$.
The timestamp of message/anti-message would therefore be missing in the global reduction while computing GVT. However, in our proposal, this timestamp gets recovered (if not yet belonging to the past of the computation) by having the worker-threads computing $min_*^B$ when we are sure that any message sent by some worker-thread up to the end of {\sf phase-A} is (or has been, if already processed) observable.

How to carry on the different phases in a wait-free manner and how to embed this GVT calculation scheme into a classical thread execution flow for Time Warp systems is discussed in the following section, where the pseudo-code of our protocol is provided.

\remove{

After the {\sf send-phase} ends, the worker-threads are requested to re-execute the incorporation

  An additional phase is interposed between

This organization leads to having that any message/anti-message in the Time Warp system is at any time incorporated into some data structure that is accessible to at least one worker-thread. In fact the send operation of any message/anti-message boils down to incorporating the sent information into the incomung messaging data structure of the destination worker-thread.

, and in selecting the next LP to be dispatched, tipically according to the next-timestamp-first algorithm.

}

\subsection{Protocol Pseudo-code}

To support the detection of the end of each phase of the protocol, we use atomic counters. The startup of the GVT protocol is therefore handled according to what reported in Algorithm \ref{init}. It simply consists in setting the $GVT\_flag$ shared variable to the value $TRUE$ after having set to the value $N$ (number of participating threads) five different atomic counters. The counters $C_A$,  $C_{send}$ and $C_B$ directly map to the above presented protocol execution phases, while the other atomic counters $C_{aware}$ and $C_{end}$ are used to identify the completion of two additional phases where the worker-threads actually become aware of the newly computed GVT value, so that the GVT protocol is allowed to terminate and to to be triggered again for a subsequent computation.

\begin{algorithm}
\caption{GVT\_INIT}
{\footnotesize
$C_A=C_{send}=C_B=C_{aware}=C_{end}=N$; // shared atomic counters\\
$GVT\_flag=TRUE$; // shared flag
}
\label{init}
\end{algorithm}

\begin{algorithm}
\caption{Main simulation loop (worker-thread $WT_i$)}
\label{main-loop}
{\small
$current\_GVT\_round\leftarrow 0$;~~~//shared round-counter\\
$my\_phase_i \leftarrow A$;~~~~~~~~~// thread local\\
$my\_GVT\_round_i \leftarrow 0$; ~~//thread local~\\
~\\
{\bf while} (not end)\{\\
1~~incorporate messages into event-queue;\\
2~~execute next-event (if any);\\
3~~send output messages/anti-messages (if any);\\

4~~switch($GVT\_flag$)\{\\
~\\
5~~~~{\bf case} FALSE:\\
6~~~~~~{\bf if}($my\_phase_i = end)$\{\\
7~~~~~~~~$my\_phase_i\leftarrow A$; //back to phase-A for next GVT round
8~~~~~~~~$atomic\_dec(C_{end})$;\\
9~~~~~~\}\\
10~~~~~{\bf break};\\
~\\
11~~~{\bf case} TRUE:~\\
12~~~~~$my\_GVT\_round_i \leftarrow current\_GVT\_round$;\\
~\\
13~~~~~{\bf if} ($my\_phase_i = A)\{$\\ 
14~~~~~~~ incorporate messages into event-queue;\\
15~~~~~~~~compute $min_i^A$;\\
16~~~~~~~~$my\_phase_i \leftarrow send$; // entering phase-send\\
17~~~~~~~~$atomic\_dec(C_A)$; // notify finalization of phase-A\\
18~~~~~~~~{\bf break};\\
19~~~~~\}\\

20~~~~~{\bf if} ($my\_phase_i = send ~\&\&~C_{A}=0)\{$\\ 
21~~~~~~~ incorporate messages into event-queue;\\
22~~~~~~~~execute next-event (if any);\\
23~~~~~~~~send output messages/anti-messages (if any);\\
24~~~~~~~~$my\_phase_i \leftarrow B$; // entering phase-B\\
25~~~~~~~~$atomic\_dec(C_{send})$; // notify finalization of phase-send\\
26~~~~~~~~{\bf break};\\
27~~~~~\}\\
~\\
28~~~~~{\bf if} ($my\_phase_i = B~ \&\&~ C_{send} = 0)\{$\\ 
29~~~~~~~ incorporate messages into event-queue;\\
30~~~~~~~~compute $min_i^B$;\\
31~~~~~~~~store $min(min_i^A,min_i^B)$;\\
32~~~~~~~~$my\_phase_i \leftarrow aware$; // entering phase-aware\\
33~~~~~~~~$atomic\_dec(C_B)$; // notify finalization of phase-B\\
34~~~~~~~~{\bf break};\\
35~~~~~\}\\

36~~~~~{\bf if} ($my\_phase_i = aware~\&\&~C_{B}=0)\{$\\ 
37~~~~~~~~compute $GVT$ as the global min of all stored local min;\\
38~~~~~~~~$my\_phase_i \leftarrow end$; // entering phase-end\\
39~~~~~~~~~$atomic\_dec(C_{aware})$; //notify finalization of phase-aware\\
40~~~~~~~~~{\bf if} ($C_{aware}$ = 0)\{\\
41~~~~~~~~~~~~$CAS(GVT\_flag,TRUE,FALSE)$;\\
42~~~~~~~~~\}\\
43~~~~~~~~{\bf break};\\
44~~~~~\}\\
~\\
\hspace*{0.4cm}~\}//end switch\\
\}//end while
}
\end{algorithm}

By the structure of Algorithm \ref{main-loop}, in case the $GVT\_flag$ is found to be set to $TRUE$, the generic worker-thread $WT_i$
immediately ends its permanence in {\sf phase-A}, which is done by computing $min_i^A$ (see line 15), updating its local phase variable to $send$, and notifying to the other worker-threads that it ended this phase (see line 17). While executing the main loop, with the local phase variable set to $send$, as soon as all the processes have ended their execution in {\sf phase-A} (namely $C_A$ is found to be zero---see line 20), the thread is forced to execute at least one event, by also sending output messages/anti-messages (if any) towards the other threads. Then it again notifies the end of the current phase (by decrementing $C_{send}$---see line 25), and sets its local phase variable to $B$. When all the threads have done the same (hence $C_{send}$  has reached the value 0), {\sf phase-B} can start. Hence each thread $WT_i$ incorporates the incoming messages into the event queue, and then computes $min_i^B$ and its local minimum, which gets then stored in memory. It also decreases $C_B$ and moves its phase to $aware$, so that when this counter reaches the value 0, all the local minima are already stored, and any worker-thread can compute the new GVT value (see lines 36--43), by also decrementing $C_{aware}$, so as to indicate awareness of the new GVT value. When this counter reaches the value 0, then the $GVT\_flag$ is reset via a compare-and-swap (CAS) atomic operation, which leads only one of the worker-treads to succeed in the reset task. At this point, the worker-threads trap into the code block in lines 5-10 hence re-initializing their phase to $A$ (so as to allow the GVT protocol to correctly restart for a subsequent round). When all of them have re-initialized their phase control-variable, the $C_{end}$ counter reaches the value 0, so that the {\bf if} condition in Algorithm \ref{trigger-init} can be satisfied. Hence the GVT protocol can be re-triggered for a subsequent round, which is done atomically via a CAS operation involving a global (shared) round-counter $current\_GVT\_round$, and a local one $local\_GVT\_round_i$. Only one worker-thread will be allowed to update the shared round-counter, also triggering the restart of the GVT protocol. On the other hand, any worker-thread re-aligns its local round-counter to the shared one as soon as it becomes aware that the GVT protocol has been started by some tread (see line 12 of Algorithm \ref{main-loop}).

With this scheme, no worker-thread enters any wait phase due to delays in the processing of specific GVT protocol steps by any other thread in the system (e.g. because of a context-switch off the CPU for any reason like, e.g., a page fault), which is were the wait-free property of our solution stands.

\begin{algorithm*}
\caption{GVT protocol start (worker-thread $WT_i$) - this might be triggered by timeout}
{\footnotesize
{\bf if} $(GVT\_flag=FALSE$ \&\& $C_{end}=0$ \&\& $CAS(current\_GVT\_round, local\_GVT\_round_i, local\_GVT\_round_i+1) )$\\
~~{\bf goto} GVT\_INIT;
}
\label{trigger-init}
\end{algorithm*}


\section{Experimental Data}
\label{data}
\subsection{Experimental Platform}
We have integrated%
\footnote{The source code of our implementation is available at \url{http://svn.dis.uniroma1.it/svn/hpdcs/root_sim/trunk/src/simulator/subsystems/gvt/}.}
the presented wait-free GVT protocol within the {\sf ROOT-Sim} simulation platform \cite{rootsim}. This is a {\sf C}-based open source simulation package targeted at POSIX systems, which implements
a general-purpose parallel/distributed simulation environment relying on the optimistic Time Warp
synchronization paradigm. It offers a very simple programming model relying on the classical notion
of simulation-event handlers (both for processing events and for accessing a committed and globally
consistent state image upon GVT calculations), to be implemented according to the ANSI-C standard,
and transparently supports all the services required to parallelize the execution. More in detail, we integrated the wait-free
GVT protocol in the symmetric multi-threaded version
of {\sf ROOT-Sim} that has been presented in \cite{Vit12b}. This version is explicitly tailored for shared-memory multi-core platforms, and offers automatic support for load-sharing and balanced distribution of the overall simulation workload across the worker-threads running within the platform. This version adheres to the observability property of Time Warp systems given that message/anti-message exchange across different worker-threads is supported by directly en-queuing the sent information within a bottom-half queue, which is accessible by the destination worker-thread at any time. On the other hand, the partitioning of the message send across top/bottom-half operations allows for high scalability of the simulation platforms.

This platform has been run on top of a 32-core HP ProLiant server equipped with 64GB of RAM and
running Debian 6 on top of the 2.6.32-5-amd64 Linux kernel.

\subsection{Test-bed Application}

In this experimental study, we have used the PHOLD benchmark application \cite{Fuj90b}. Particularly, we have run the specific
implementation of this benchmark presented in \cite{Vit09}. It is based on homogeneous LPs that perform memory allocation/deallocation operations (across lists of buffers of different sizes) as well as memory read/write operations onto these buffers. The events occurring in the system lead the LPs to modify their memory layout (beyond performing read/write operations). In the configuration of the benchmark that we have used, each LP schedules for itself memory-deallocation events that, once processed, lead the LP to deallocate some buffers belonging to its state. On the other hand, when de-allocating some buffer, a memory-allocation event is scheduled for some other LP in the system. The concept underlying this implementation/configuration of the PHOLD benchmark is to have a stable value of the global amount of memory used to represent the state of the whole simulation model, by also having continuous variations of the amount of memory used for the local state of individual LPs. Also, scheduled simulation events tend to be clusters (along simulation time) across LPs exhibiting higher memory usage for their states. This is useful in our tests given that we may have short-lived execution phases where a few worker-threads may have few events to process, which possibly increases the likelihood for these threads to compete for the access to critical sections (if any) used to support specific housekeeping operations, such as GVT computation (which is especially true when fixing the size of the simulation model, and running with larger numbers of worker-threads). This is therefore a good test case for evaluating wait-free implementations of Time Warp-suited housekeeping protocols like the GVT protocol we are presenting.

\subsection{Results}

We have compared the run-time behavior of our wait-free GVT protocol (which we refer to as WF in the plots) with the one of the algorithm by Fujimoto and Hybinette (referred to as FH). The latter has been also integrated within {\sf ROOT-Sim} as an alternative to WF. Particularly, we have fixed the size of the used PHOLD model to 32 LPs (with a total amount of live memory for the corresponding states of the order of  1 GB --- about 32 MB per-LP on the average), with read (resp. write) operations touching 20\% (resp. 10\%) of the current LP state size, and we have performed experiments in two different configurations of number of worker-threads within the simulation platform and workload conditions on the underlying computing platform.

\begin{figure}[t]
\centering
\includegraphics[height=1.8in]{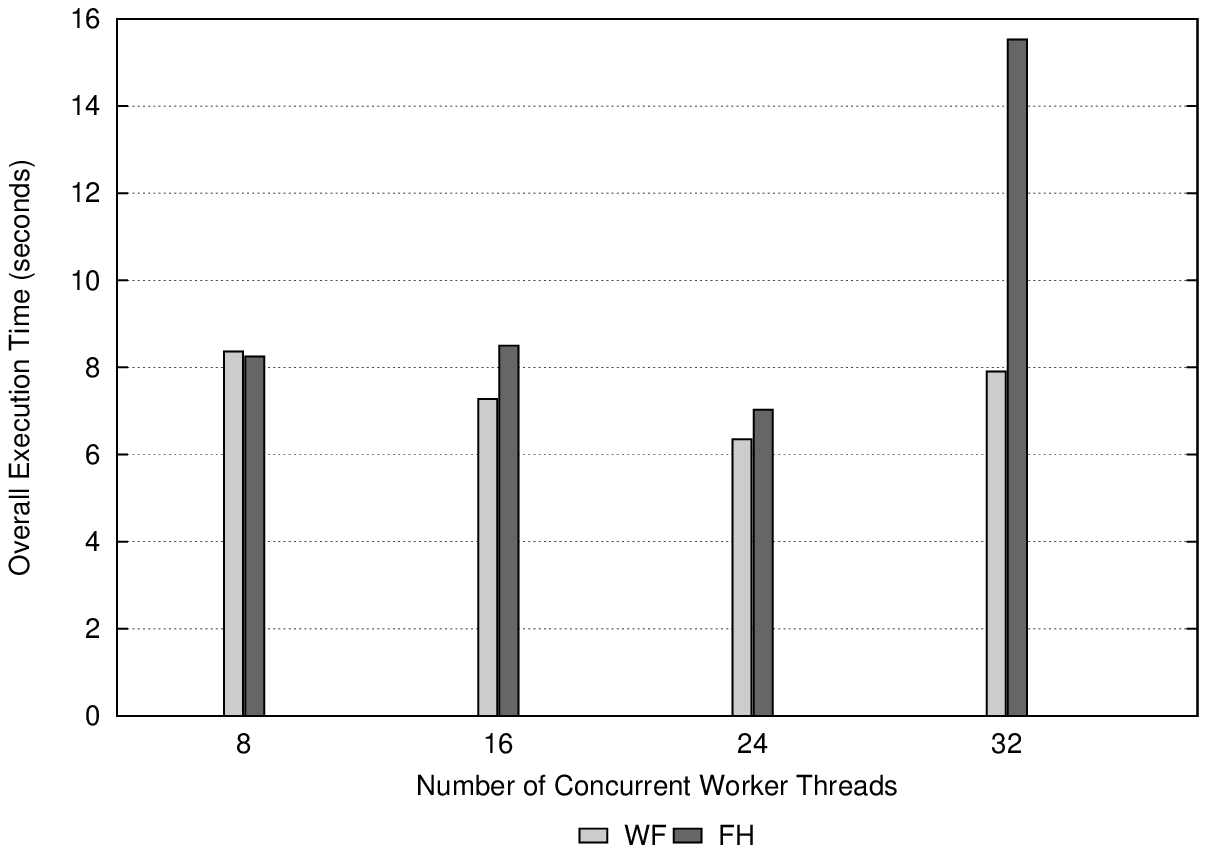}
\includegraphics[height=1.8in]{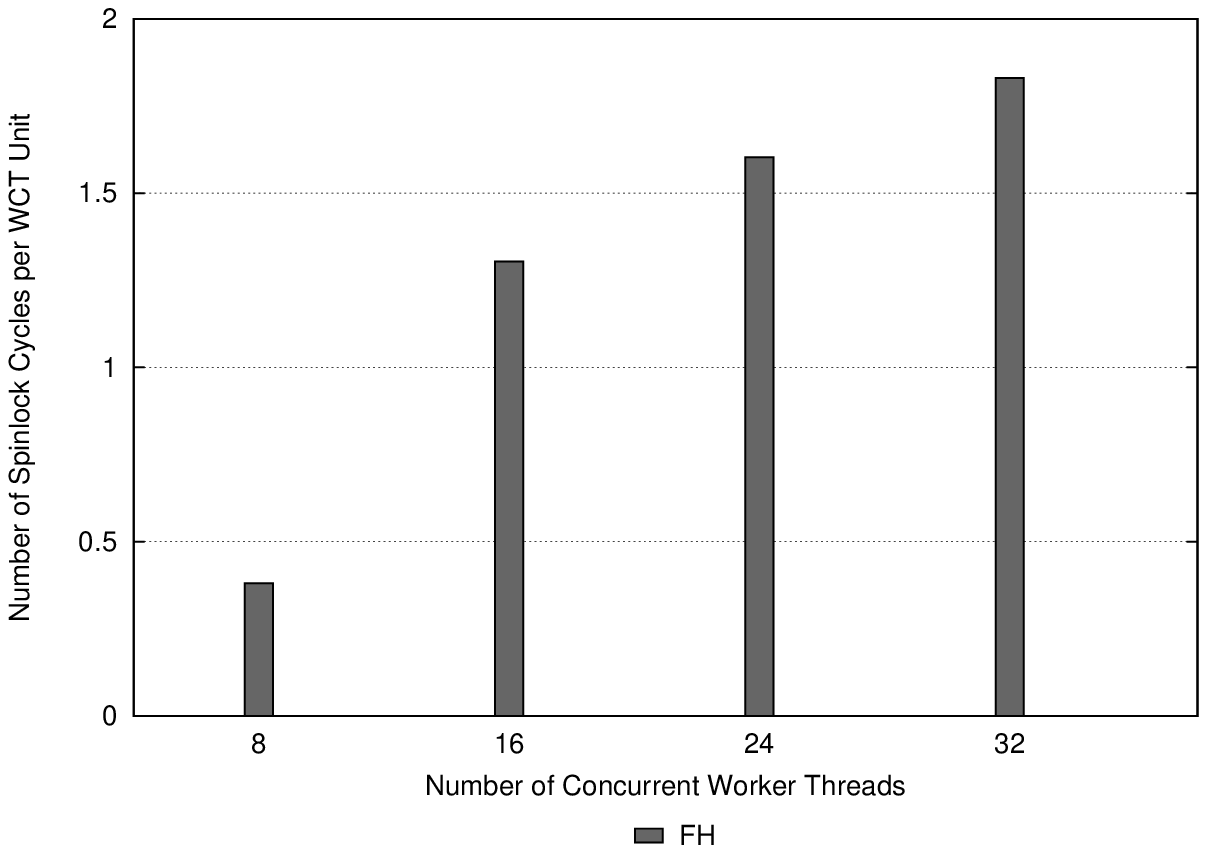}
\caption{Results with the computing platform reserved for the simulation runs.}
\label{config-1}
\end{figure}

In the first configuration,  the computing platform has been reserved for the simulation runs, and we have varied the number of worker-threads used for running the PHOLD application between 2 and the maximum number of available hardware-cores, namely 32.
We recall that, having the size of the PHOLD model been fixed to 32 LPs, variation of the number of worker-threads towards the maximum value of 32 leads to scenarios with increasing parallelism, and hence increased likelihood of the aforementioned phenomenon where, for short-lived phases, some worker-thread may not have simulation work to be performed (being the scheduled events temporarily clustered to occur  on LPs hosted by other worker-threads, just depending on how the size of the memory used to keep  the LPs' states varies over time depending on deallocation/allocation operations occurring upon processing the events).
For this experiment, we fixed the timeout for triggering a new GVT computation to 1 sec (this value looks reasonable for allowing prompt recovery of memory, while not making GVT computation a predominant housekeeping task), and we measured the wall-clock-time required to complete the run of the PHOLD model in the selected configuration. The reported values refer to the average wall-clock-time observed over 10 different runs, all done with different pseudo-random seeds. However, the same seed is used for the corresponding runs with the two different GVT protocols so as to allow the same trajectory for the evolution of the simulation model when taking each individual wall-clock-time sample for the two protocols. For the case of FH, we also report the number of spin-lock tries per wall-clock-time unit experienced by the worker-threads while attempting to enter the critical section proper of this GVT protocol, which provides an indication of how the worker-threads tend to compete in the access to GVT-support structures of FH, in mutual exclusion, while increasing the level of parallelism.
The outcoming results are shown in Figure \ref{config-1}, and we observe that, while increasing the number of worker-threads, the WF protocol allows for reducing the wall-clock-time required to complete the run up to 50\% when compared to FH, a phenomenon which is strictly related to the higher overhead paid by FH in terms of CPU-time requested for running tasks related to GVT computation. In fact, as shown by the data on the bottom of Figure \ref{config-1}, as soon as the degree of parallelism gets increased (and the aforementioned short-lived phases with no simulation event to be processed at some worker-thread materialize),  the likelihood  of concurrent execution of the worker-threads in housekeeping mode increases, with consequent increase of the incidence of the critical-section access delay when running the FH GVT-protocol. This is avoided by WF due to its wait-free nature.

We have also measured the execution time required to run the same identical PHOLD model (same code) in a serial fashion, by relying
on a calendar-queue scheduler for storing the events and selecting the next-to-be processed one. This was on the order of 62 sec, which leads to the observation that the parallel runs provide speedup while increasing the number of worker-threads (up to about 10, which is achieved when running with 24 worker-threads and WF as the GVT protocol). Overall, the results presented in this study refer to the case of competitive parallel executions, which further strengthen the relevance of the improvement provided by our wait-free GVT protocol.

On the other hand, we note that the wall-clock-time does not scale down when running with more than 24 worker-threads, which is due to two main reasons. One is that rollbacks tend to increase while increasing the level of parallelism, thus increasing the likelihood of performing non-useful work. Second, due to the specific PHOLD configuration used, the short-lived phases with no event to be processed at some worker-thread (which especially materialize when increasing the number of worker-threads) lead to non-full exploitation of the computing power offered by the underlying platform. However, by the data we see a much higher resilience of the WF protocol towards performance degradation phenomena caused by these ``over-parallelism" scenarios.

\begin{figure}[t]
\centering
\includegraphics[height=1.8in]{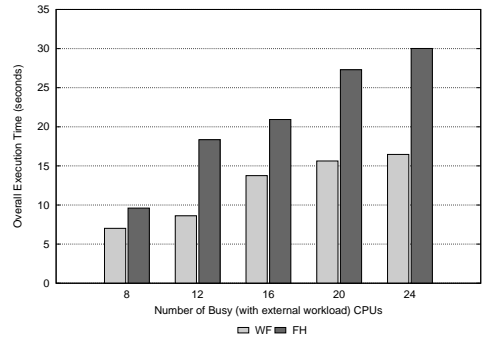}
\caption{Results with the external interfering workload.}
\label{config-2}
\end{figure}

\remove{
We also note that the wall-clock-time does not scale down linearly vs the number of worker-threads, which is due to two main reasons. One is that rollbacks tend to increase while increasing the level of parallelism, thus increasing the likelihood of performing non-useful work. Second, due to the specific PHOLD configuration used, the short-lived phases with no event to be processed at some worker-thread (which especially materialize when increasing the number of worker-threads) lead to non-full exploitation of the computing power offered by the underlying platform. Overall, the non-linear decrease of the wall-clock-time is not due to scalability problems in the used platform, rather to the specific nature of the simulation workload we are using, which simply leads to increasing the burst of worker-thread execution of its main loop in housekeeping mode. As already mentioned, this represents a good test case for assessing the efficiency of optimized housekeeping protocols, especially those requiring a form of coordination across the different worker-threads, which may amplify the actual cost of non-optimized protocol tasks.
}

In the second configuration, we ran the same PHOLD model by fixing the number of worker-threads to 24 (which was the best parallelism level observed in the previous configuration). However, we injected external workload on the computing platform, which is made up by CPU-bound processes simply executing a busy loop. The number of such processes has been varied between 8 and 24. In this scenario, we have that {\sf ROOT-Sim} worker-threads compete for CPU usage with the external workload (at least when the number of interfering CPU-bound processes oversteps the value 8), a phenomenon which can give rise to delays in the completion of the critical section characterizing FH in case the worker-thread running the critical section is context-switched off the CPU, and to consequent delays in the access to the same critical section by the other worker-threads. This experiment is therefore intended as a means to assess the resilience of WF (vs FH) to performance degradation in case of interference by external workload on the execution of the simulation platform. By the data shown in Figure \ref{config-2}, we see that the FH protocol rapidly tends to degrade the performance of the Time Warp system as soon as the external workload tends to increase. This phenomenon is noted also for WF. However, with this protocol, the degradation is mainly expected to be generated by the reduced computing power exploited by the Time Warp system and by secondary costs, such as (a) those related to cache invalidation and refill in case of context-switch between a {\sf ROOT-Sim} worker-thread and an external interfering process, and (b) those related to the increase of rollback due to more skewed advancement in logical time of the different worker-threads in case of interference. In any case, WF still provides   30\%-50\% reduction of the wall-clock-time to complete the simulation run when compared to FH, a gain which is noted as soon as minimal interference by the external workload takes place.

\section{Conclusions}

In this paper we have shown what are the implications on performance of wait-free algorithms when dealing
with coordination phases in share-memory/multi-core environments for high performance computing.
By the results, we can see that the adoption of wait-free coordination algorithms can enable for
higher performance and increased scalability, with respect to the traditional lock-based synchronization.

\end{document}